\documentclass[a4paper, 12 pt] {article}

\usepackage{amsmath}
\usepackage{amsfonts}
\usepackage{amsthm}
\usepackage{amscd} 

\usepackage{graphicx}
\usepackage{epsfig}
\usepackage{graphics}

\newcommand\Div{ \,\text{div}\, } 
\newcommand\Curl{\,\text{\bf curl}\,}
\renewcommand\Re{\,\text{Re}\,}
\renewcommand\Im{\,\text{Im}\,}

\newcommand{\half}{{\textstyle \frac{1}{2}}}
\newcommand{\fourth}{{\textstyle \frac{1}{4}}}

\newcommand{\hhalf}{{\scriptstyle \frac{1}{2}}}
\newcommand{\tthird}{{\scriptstyle \frac{1}{3}}}
\newcommand{\ffourth}{{\scriptstyle \frac{1}{4}}}
\newcommand\smallfrac[2]{{\textstyle \frac{#1}{#2}}}
\newcommand{\Dvec}{{\bf D}}
\newcommand{\nvec}{{\bf n}}
\newcommand\grad{{\bf \nabla}}
\newcommand\avec{{\bf a}}

\newcommand\cvec{{\bf c}}

\newcommand\tvec{{\bf t}}
\newcommand\uvec{{\bf u}}
\newcommand\vvec{{\bf v}}
\newcommand\wvec{{\bf w}}

\newcommand\rhat{{\bf \hat r}}
\newcommand\xhat{{\bf \hat x}}
\newcommand\sgn{\,{\hbox{\rm sgn}}}
\newcommand{\rvec}{{\bf r}}
\newcommand{\Acal}{{\cal A}}
\newcommand{\Rr}{{\mathbb R}}
\newcommand{\Cc}{{\mathbb C}}

\newcommand\octant{R} %decide on notation for octant of prism

\newcommand\fbar{{\bar f}}
\newcommand\tbar{{\bar t}}
\newcommand\wbar{{\bar w}}

\begin{document}

\title{Elastic energy of liquid crystals in convex polyhedra}

\author{A Majumdar$^{\dag\ \ddag}$, 
%\thanks{e-mail address: {\tt a.majumdar@bristol.ac.uk}} \
JM Robbins$^\dag$
%\thanks{e-mail address: {\tt j.robbins@bristol.ac.uk}} \
\& M Zyskin$^\dag$ 
\thanks{a.majumdar@bristol.ac.uk, j.robbins@bristol.ac.uk, m.zyskin@bristol.ac.uk}\\
School of Mathematics\\ University of Bristol, University Walk, 
Bristol BS8 1TW, UK\\
and\\
Hewlett-Packard Laboratories,\\
 Filton Road, Stoke Gifford, Bristol BS12 6QZ, UK}

\thispagestyle{empty}

\maketitle

\begin{abstract}
  We consider nematic liquid crystals in a bounded, convex polyhedron
  described by a director field $\nvec
(\rvec)$ subject to tangent
  boundary conditions.  We derive lower bounds for the one-constant
  elastic energy in terms of topological invariants.  For a right
  rectangular prism and a large class of topologies, we derive upper
  bounds by introducing test configurations constructed from local
  conformal solutions of the Euler-Lagrange equation.  The ratio of
  the upper and lower bounds depends only on the aspect ratios of
  the prism.  As the aspect ratio is varied, the minimum-energy 
  conformal state undergoes a sharp transition from being smooth to
  having singularities on the edges.

\end{abstract}

%\pacs{61.30.Jf,  11.10.Lm, 61.30.Dk, 61.30.Hn, 11.27.+d}

\newpage

The continuum theory of liquid crystals \cite{degennes} is a
prototypical nonlinear field theory in which topological
considerations play a fundamental role \cite{mermin, kleman,
  lavrentovich}, both in equilibrium (eg \cite{brezis, kamien2003})
and dynamical phenomena (eg \cite{yeomans}).  Nematic liquid crystals
are represented by a director field $\nvec(\rvec)$, which describes
the mean local orientation of the constituent rod-like molecules.  In
confined geometries, boundary conditions, which depend on the
properties of the substrate, can play a significant role (see, eg
\cite{patricio2002}).

In this Letter, we consider director fields in a bounded,
three-dimensional convex polyhedron $P$.  While natural theoretically,
the problem also has a technological motivation; polyhedral geometries
have been proposed as a mechanism for engendering bistability in
liquid crystal display cells \cite{newtonspiller}, \cite{kg2002}.  Polyhedral cells
can support two (or more) energetically stable director configurations
with contrasting optical properties.  Power is required only to
switch pixel states but not to maintain them. Bistable cells offer the
promise of displays with higher resolution and requiring less power
than is available with current technologies based on monostable cells.

We suppose $P$ has strong azimuthal anchoring, so that $\nvec$
satisfies {\it tangent boundary conditions} -- on the faces of $P$,
$\nvec$ must be tangent to the faces.  Tangent boundary conditions
imply that, on the edges of $P$, $\nvec$ is parallel to the edges,
and, therefore, is necessarily discontinuous at the vertices.  We
restrict our study to configurations which are continuous everywhere
else, ie as continuous as possible.  As $P$ is simply connected, we
can then regard $\nvec(\rvec)$ as a unit-vector field, rather than a
director field.  We obtain lower bounds for the elastic energy of
$\nvec$ in terms of its topological invariants, and, for a rectangular
prism, upper bounds, which differ from the lower bounds only by a
geometry-dependent factor.  The upper bounds are obtained from local
conformal solutions of the Euler-Lagrange equation, whose energetics
indicate the onset of edge singularities as the prism becomes cubic.

It turns out that tangent boundary conditions 
produce a large family of topologically
distinct configurations.  A complete classification is given in
Ref.~\cite{rz2003}, whose results we briefly summarise.  Tangent
unit-vector fields on $P$ can be classified up to homotopy (ie,
continuous deformations) by a family of invariants: the {\it edge
  orientations}, {\it kink numbers} and {\it trapped areas}.  The edge
orientations are just the values of $\nvec$ on the edges of $P$, and
therefore are essentially signs.  The kink numbers determine the
number of times $\nvec$ winds along an (outward-oriented) path on a
face of $P$ which joins a pair of adjacent edges.  On such a path, the
initial and final values of $\nvec$ are determined by the edge
orientations; in between, $\nvec$ describes a curve on the circle of
unit vectors tangent to the face.  The shortest curve between the
endpoints is assigned kink number 0.  In general, the kink numbers are
integers.  The trapped areas, $\Omega^a$, are defined as follows.  Let
$C^a$ be a surface inside $P$ which
%CHANGE
separates the vertex $a$ from the other vertices.  Then $\Omega^a$ is the area of the
region $\nvec(C^a)$ on the unit two-sphere.  This may be written
as $\int_{C^a} \Dvec \cdot\,{\bf dS}$, where ${\bf dS}$ is
the outward-oriented area element on $C^a$, and
\begin{equation}
  \label{eq:D}
%  \Dvec = \epsilon_{\alpha\beta\gamma} n_\alpha
%(\nabla n_\beta \times \nabla_\gamma \nvec).
D_j = \half 
\epsilon_{jkl} (\partial_k \nvec \times \partial_l \nvec)\cdot \nvec.
\end{equation}
$\Dvec$ may be regarded as the vector field dual to the pull-back,
$\nvec^*\omega$, of the outward-oriented area form $\omega$ on the
sphere (in polar coordinates, $\omega = \sin\theta\,d\theta\wedge
d\phi$).  $\Omega^a$ is independent of the choice of $C^a$, and may be
thought of as the degree of a fractional point defect at $a$.
$\Omega^a$ need not be integral, but the allowed values of $\Omega^a$
for given edge orientations and kink numbers differ by integer
multiples of $4\pi$ (complete coverings of the sphere).  These
integers are the wrapping numbers.  The invariants satisfy certain sum
rules, which follow from the fact that $\nvec$ is continuous away from
the vertices.  The sum of the kink numbers on each face is determined
by the edge orientations, while the trapped areas sum to zero.
All values of the invariants consistent with the sum rules can be
realised, and two configurations are homotopic if and only if their
invariants are the same.

%3. Lower bound

In the continuum theory, the energy of $\nvec$
is given by the elastic, or Frank-Oseen, energy \cite{degennes},
%% \begin{multline}
%%   \label{eq:elastic}
%%   E = \int_P \big [ K_1 (\Div \nvec)^2 + K_2 (\nvec
%%   \cdot \Curl \nvec)^2 + \\  K_3
%%   (\nvec\times \Curl \nvec)^2 +  K_4 \Div ((\nvec\cdot
%%   \grad)\nvec - (\Div \nvec) \nvec)\big ]\, dV.
%% \end{multline}
\begin{multline}
  \label{eq:elastic}
  E = \int_P \big [ K_1 (\Div \nvec)^2 + K_2 (\nvec
  \cdot \Curl \nvec)^2 +   K_3
  (\nvec\times \Curl \nvec)^2\\ +  K_4 \Div ((\nvec\cdot
  \grad)\nvec - (\Div \nvec) \nvec)\big ]\, dV.
\end{multline}
Tangent boundary conditions 
%TBCs 
imply that the contribution from the 
$K_4$-term, which is a pure divergence, vanishes.  In the so-called
one-constant approximation, the remaining elastic constants $K_1$,
$K_2$ and $K_3$ are taken to be the same.  In this case,
(\ref{eq:elastic}) simplifies to
%CHANGE
\begin{equation}
  \label{eq:one-constant}
  E = K \int_P  (\grad \nvec)^2\,  dV = 
K \int_P  \sum_{j=1}^3 (\partial_j \nvec)^2  \, dV.
\end{equation}
Stable (or quasi-stable) configurations are minima (or local minima)
of the energy.  In the one-constant approximation, local minimisers
satisfy the Euler-Lagrange equation and boundary conditions 
 \begin{gather}
   \label{eq:EL}
   \Delta \nvec = (\nvec\cdot\Delta\nvec) \nvec,\\ 
(\cvec \cdot
   \nvec)_{\partial P} = 
   (\cvec\times\grad_c\nvec)_{\partial P}  = 0,\nonumber
 \end{gather}
%\begin{equation}
%  \label{eq:EL}
%  \Delta \nvec = (\nvec\cdot\Delta\nvec) \nvec,\quad (\cvec \cdot
%  \nvec)_{\partial P} = 
%  (\cvec\times\grad_c\nvec)_{\partial P}  = 0,
%\end{equation}
where
$\cvec$ is the outward unit-normal on the boundary $\partial P$ 
of $P$, and $\grad_c$ denotes the
derivative along $\cvec$ (the normal derivative here);
the condition on $\grad_c\nvec$ ensures there is no
boundary contribution to the first-order variation of the elastic energy.  
(Minimisers of the energy (\ref{eq:one-constant})
are called
{\it harmonic maps} in the
mathematics literature.)

Our first result is a lower bound for the energy
\eqref{eq:one-constant} 
in each homotopy
class.  
%The lower bound turns out to depend only the trapped areas
%$\Omega^a)$.  
%CHANGE
The derivation extends an argument from a seminal paper of Brezis, Coron and Lieb
\cite{brezis} to the polyhedral boundary-value problem we are 
considering here.
Below we restrict
to the case where $\grad \nvec$ is smooth away from the vertices.
%(say piecewise continuously differentiable).  
In fact, the bound
is good provided only that $\grad \nvec$ is square-integrable \cite{mrz2004a}.

Since $\nvec$ is normalised, the partial
derivatives $\partial_j \nvec$, $j = 1, 2, 3$, are orthogonal to
$\nvec$, and therefore are linearly dependent.  Thus, at each point
$\rvec$, there is at least one direction, $\wvec$ say, in which the
derivative of $\nvec$ vanishes.  Let us introduce unit vectors $\uvec$
and $\vvec$ which together with $\wvec$ define an orthonormal frame.
Then 
\begin{equation}
  \label{eq:first inequality}
  (\grad\nvec)^2 = (\grad_u \nvec)^2 + (\grad_v \nvec)^2
  \ge 2 |\grad_u \nvec \times \grad_v \nvec|,
\end{equation}
where equality holds if and only if $\grad_u\nvec$ and $\grad_v\nvec$
are orthogonal and of equal length.  From (\ref{eq:D}), the right-hand side
of (\ref{eq:first inequality}) is just $2|(\uvec\times
\vvec)\cdot \Dvec|$, which in turn is just equal to $2 |\Dvec|$, since,
by inspection, $\Dvec$ is orthogonal to $\uvec$ and $\vvec$.  Then
%It follows that
\begin{equation}
  \label{eq:grad n^2}
  (\grad\nvec)^2 \ge  2|\Dvec|.
\end{equation}

Let $\xi(\rvec)$ be a continuous
scalar function with piecewise continuous gradient such that $|\grad
\xi| = 1$.  Then $|\Dvec| \ge \Dvec\cdot\grad\xi$.  Let $\hat P$ be a
domain obtained by excising from $P$ vanishingly small neighborhoods of
each vertex.  From
(\ref{eq:one-constant}) and (\ref{eq:grad n^2}), it follows that
\begin{equation}
  \label{eq:energy >=}
  E \ge 2K \int_{\hat P} \Dvec\cdot \grad \xi\,dV.
\end{equation}
%$ E \ge 2K \int_{\hat P} \Dvec\cdot \grad \xi\,dV$.
But $\Dvec$ has zero divergence, as follows from the fact that 
area-form $\omega$
is closed (as well as from direct calculation).
Integrating by parts, we get 
\begin{equation}
  \label{eq:E >= 2nd} 
  E \ge 2 K \oint_{\partial \hat P} \xi \Dvec\cdot{\bf dS}.
\end{equation}
%$ E \ge 2 K \oint_{\partial \hat P} \xi \Dvec\cdot{\bf dS}$.  
The boundary of $\hat P$, denoted $\partial \hat P$, consists of (parts of) the
faces of $P$ as well as the boundaries of the excised regions.  
But $\Dvec$ vanishes on the faces of $\hat P$, as 
tangent boundary conditions
%TBCs 
imply that the tangential
derivatives of $\nvec$ are parallel to each other there.
On the boundaries of the excised
regions, $\xi$ can be set to its values $\xi^a$ at the vertices.
Thus, the
integral $\int_{\partial \hat P} \xi \Dvec\cdot\, {\bf dS}$ reduces to a weighted
sum of fluxes of $\Dvec$ through the excised boundaries.  These fluxes
are just the trapped
areas.  We obtain the lower bound
 \begin{equation} 
   \label{eq:E >= 2nd}
 E \ge 2 K \sum_a \xi^a \Omega^a.
 \end{equation}
%$E \ge 2 K \sum_a \xi^a \Omega^a$.  
To proceed, note the values $\xi^a$ are
constrained; since $|\grad\xi| = 1$, we must have $|\xi^a - \xi^{b}|
\le L^{ab}$, where $L^{ab}$ is the length of the edge between vertices
$a$ and $b$.  Conversely, given a set of values $\xi^a$ satisfying
these inequalities, we can construct a function $\xi(\rvec)$ which
interpolates between these values, and for which $\grad\xi$ is piecewise
continuous with unit norm (for example, let $\xi(\rvec) = \max_a
(\xi^a - |\rvec-\avec|)$).  Therefore, we can express the
lower bound (\ref{eq:E >= 2nd}) in terms of the trapped areas and the lengths of
the edges of $P$,
\begin{equation}
  \label{eq:lower bound}
  E \ge 2K \max_{\xi^a:  
|\xi^a - \xi^{b}| \le L^{ab}} 
\sum_a \xi^a  \Omega^a.
\end{equation}
(\ref{eq:lower bound}) specifies a linear optimisation problem, which
can be solved by standard methods.  

We note that because $\sum_{a}
\Omega^a = 0$, the optimal solution $\xi^a$ is in general only
determined up to an overall additive constant.  If the one-constant
approximation is dropped, $K$ in \eqref{eq:lower bound} can be
replaced by the smallest of the three elastic constants $K_1$, $K_2$
and $K_3$.

Next, we consider the case where 
$P$ is a right rectangular prism, with sides of
length $L_x \ge L_y \ge L_z$.  Moreover, within $P$, we consider
configurations (and their homotopy classes) 
which are {\it
  reflection-symmetric} about the mid-planes of $P$.  That is,
 \begin{equation}
   \label{eq:reflection symmetric}
   \nvec(L_x - x,y,z) = 
   \nvec(x,L_y - y,z) =
   \nvec(x,y,L_z-z) = \nvec(x,y,z)
 \end{equation}
%$\nvec(L_x - x,y,z) = \nvec(x,y,z)$, 
%$\nvec(x,L_y - y,z) = \nvec(x,y,z)$, etc
%$\nvec(x,y,L_z-z) = \nvec(x,y,z)$.
(we plan to treat more general geometries and homotopy classes in the
future).  Clearly, a reflection-symmetric configuration is determined
by its values in the octant of the prism, denoted $\octant = \{0\le r_j \le
\half L_j\}$, as are its
invariants.  In particular, the trapped areas at two vertices are
either the same or differ by a sign according to whether the vertices
are related by an even or odd number of reflections.  
Let $\Omega^0$ denote the trapped area at the origin.
Taking $\xi^a$ in (\ref{eq:lower bound}) (optimally) to be 
$L_z$ or $0$ according
to whether $\Omega^a$  is $|\Omega^0|$ or  $-|\Omega^0|$,
we obtain the
explicit lower bound $E \ge E_-$, where
%CHANGE
\begin{equation}
  \label{eq:LB cube}
  E_- =  8 KL_z |\Omega^0|.
\end{equation}

In the remainder of this Letter, we introduce a family of
reflection-symmetric configurations which are local solutions of the
Euler-Lagrange equation (\ref{eq:EL}). From these we infer upper
bounds for the energy as well as the onset of edge
singularities as the geometry is varied.

%% Candidates for low-energy configurations are those for which
%% \eqref{eq:grad n^2} becomes an equality.  This is the case if, on
%% $\octant$, $\nvec$ is radially constant
%% %about the origin
%% (ie, $\nvec(\lambda \rvec) = \nvec(\rvec)$) and conformal.  Here conformal
%% means that the map $\tvec\mapsto \grad_t \nvec(\rvec)$ from vectors
%% $\tvec$  orthogonal to $\rhat$ 
%% into vectors $\grad_t \nvec(\rvec)$ orthogonal to $\nvec(\rvec)$, 
%% preserves orientation, angles and ratios of
%% lengths (or else vanishes).  If $\nvec$ is radially constant then $\Dvec$ is radial (cf
%% \eqref{eq:D}), while if $\nvec$ is also
%% conformal, then $\Dvec = \half (\grad \nvec)^2\rhat$.  %CHANGE
%% It is straightforward to verify that radially constant, conformal
%% configurations satisfy (\ref{eq:EL}) in $\octant$ (but not
%% on the mid-planes of $P$, where the normal
%% derivatives of $\nvec$ are discontinuous).

Candidates for low-energy configurations are those for which
\eqref{eq:grad n^2} becomes an equality.  This is so if, on
$\octant$, $\nvec$ is radially constant
%about the origin
(ie, $\nvec(\lambda \rvec) = \nvec(\rvec)$) and conformal.  Here, conformal
means that the map $\tvec\mapsto \grad_t \nvec(\rvec)$ from vectors
$\tvec$  orthogonal to $\rhat$ 
to vectors $\grad_t \nvec(\rvec)$ orthogonal to $\nvec(\rvec)$
preserves orientation, angles and ratios of
lengths (or else vanishes).  If $\nvec$ is radially constant then $\Dvec$
is radial
(cf \eqref{eq:D}), 
while if $\nvec$ is also
conformal, $\Dvec = \half (\grad \nvec)^2\rhat$.  %CHANGE
It is straightforward to verify that radially constant, conformal
configurations satisfy (\ref{eq:EL}) in $\octant$ (but fail to
on the mid-planes of $P$, where the normal
derivatives of $\nvec$ are discontinuous).

Suppose $\nvec$ is a reflection-symmetric configuration which is
radially constant and conformal in $\octant$ (below we will show how to
construct such configurations).  Then, from
\eqref{eq:one-constant}, and using reflection symmetry, we get 
\begin{equation}
  \label{eq:upper 1}
  E = 8 K \int_{\octant}  (\grad \nvec)^2 dV =
16 K \int_{\octant} \rhat\cdot \Dvec  dV = 
16 K  \int_{\partial \octant} r \Dvec \cdot {\bf dS},
\end{equation}
where we have used $\grad\cdot \Dvec$ = 0.  We obtain an upper bound $E_+$
for $E$ by replacing $r$ by its maximum value on $\partial \octant$, namely
$\half(L_x^2 + L_y^2 + L_z^2)^{1/2}$.  The integral which remains is just
the flux of $\Dvec$ through $\octant$.  As 
tangent boundary conditions
%TBCs 
imply there is no contribution from
the exterior faces of $\octant$ (ie, $x = 0$, $y=0$ or $z=0$), 
%(tangent boundary conditions), 
this just gives
the trapped area.  Thus 
%CHANGE
\begin{equation}
  \label{eq:upper final}
  E_+ = 8 K (L_x^2 + L_y^2 + L_z^2)^{1/2}|\Omega^0|.
\end{equation}
Like $E_-$,  $E_+$
is proportional to $|\Omega^0|$, and $E_+/E_- =
(a_{xz}^2 + a_{yz}^2 + 1)^{1/2}$, where $a_{ij} = L_i/L_j$ denote
the aspect ratios of the prism.  For the cube, this ratio is $\sqrt{3}$.

It remains to construct radially constant, conformal configurations  
$\nvec(\rvec)$ in $\octant$
satisfying 
tangent boundary conditions.  
%TBCs.  
For this, it is convenient to introduce the stereographic
projection $(e_x,e_y,e_z)\mapsto (e_x + i e_y)/(1 + e_z)$ from the
sphere to the extended complex plane $\Cc^e$.  Projecting both $\rvec$
and $\nvec$, we obtain a complex-valued function $f(w)$ of complex
argument $w$ given by
\begin{equation}
  \label{eq:stereograph}
  \left(\frac{n_x + in_y}{1 + n_z}
  \right)(x,y,z) = 
f\left(
\frac{x + iy}{r + z}
\right).
\end{equation}
The
domain of $f$ is the stereographic image of $\octant$, which is the
quarter-disc $Q$ given by $|w| \le 1$ and $\Re w, \Im w \ge 0$.  
$\nvec$ conformal is equivalent to $f$ locally analytic. 
%In
%what follows, we
%consider configurations for which $f$ has a meromorphic extension from
%$Q$ to
%$\Cc^e$.

The form of $f$ is determined by
tangent boundary conditions.
%TBCs.  
Under stereographic projection, the $xz$-face is mapped to the
real $w$-axis, so we require that i) $f(w)$ is real if $w$ is real.
Similarly, the $yz$- and $xy$-faces are mapped to the imaginary axis
and the unit circle respectively, so we require that ii) $f(w)$
is imaginary if $w$ is imaginary, and iii) $|f(w)| = 1$ if $|w| = 1$.  
%%
%%
%% To proceed, we assume that $f$ has a meromorphic extension from $Q$
%% to $\Cc^e$.  Then i)--iii) may be continued to the following
%% functional equations: (a) $\fbar(w) = f(w)$, (b) $\fbar(-w) = -f(w)$,
%% and (c) $f(w)\fbar(1/w) = 1$.
%%
%% Under stereographic projection, the $xz$-plane is mapped to the
%% real $w$-axis, so we require that 
%% \begin{subequations}\label{eq: complex bcs}
%% \begin{align}
%%   &f(w)\text{\ is real} \text{\ if $w$ is real,}\label{eq:i)}\\
%%   &f(w)\text{\ is imaginary}\text{\ if $w$ is imaginary,}\label{eq:ii)}\\
%%   &|f(w)| = 1 \text{\ if\ } |w| = 1.\label{eq:iii)}
%% \end{align}
%% \end{subequations}
%% i) $f(w)$ is real if $w$ is real.
%% Similarly, the $yz$- and $xy$-planes are mapped to the imaginary axis
%% and the unit circle respectively, so we require that ii) $f(w)$ is
%% imaginary if $w$ is imaginary, and iii) $|f(w)| = 1$ if $|w| = 1$.  
To proceed, we assume that $f$ has a meromorphic extension from $Q$ to
$\Cc^e$.  Then the conditions i)--iii) may be continued
to the following functional equations:
\begin{subequations}\label{eq: funct eqns}
\begin{align}
  \fbar(w) &= f(w),\label{eq:a)}\\
  \fbar(-w) &= -f(w),\label{eq:b)}\\
f(w) \fbar(1/w) &= 1.
\label{eq:c)}
\end{align}
\end{subequations}
Together, (\ref{eq:a)}) and (\ref{eq:c)}) imply that $f(w)f(1/w) = 1$, which in turn implies
that if $w$ is a zero of $f$, then $1/w$ is a pole.  $f$ meromorphic
implies the number of zeros and poles in the unit disk must be finite
(otherwise there would be an accumulation of one or the other).  
It follows that $f$ has
only a finite number of zeros and poles in $\Cc^e$,
so that $f$ is rational.

(\ref{eq:b)}) and (\ref{eq:c)}) imply
that, if $w$ is a zero of $f$, then so are $\wbar$ and $-w$, and
similarly if $w$ is a pole.  One can then show that every reciprocal pair
$(w,1/w)$ of zeros and poles has a unique representative in $Q$.  
%This
%leads to the following parameterisation of $f$, in which, 
For
convenience, we divide these representatives into three groups: 
the strictly real
$r_1,\ldots, r_a$, with $0 < r_j < 1$  (as (\ref{eq:c)}) rules out
zeros and poles of modulus one); the strictly imaginary
$i s_1,\ldots, i s_b$, with $0 < s_k < 1$; and the strictly complex
$ t_1,\ldots, t_c$, with $0 < |t_l| < 1$.  In addition, $w = 0$ is a
zero or pole of odd order, as is implied by
(\ref{eq:a)}) and (\ref{eq:b)}).
Then $f$ is given by
%\begin{widetext}
%%   \begin{equation}
%%     \label{eq:w(z)}
%%     f(w) = \epsilon w^n 
%%   \prod_{j=1}^a\left(
%%   \frac{w^2 - r_j^2}{r_j^2 w^2 - 1}
%%   \right)^{\rho_j}
%%   \prod_{k=1}^b\left(
%%   \frac{w^2 + s_k^2}{s_k^2 w^2 + 1}
%%   \right)^{\sigma_k}
%%   \prod_{l=1}^c\left(
%%   \frac{w^2 - t_l^2)(w^2 - { t^*}_l^2)}
%%   {(t^2_l w^2 - 1)({t^*}_l^2w  - 1)}
%%   \right)^{\tau_l}.
%%   \end{equation}
%\end{widetext}
 \begin{multline}
    \label{eq:w(z)}
    f(w) = \epsilon w^n 
  \prod_{j=1}^a\left(
  \frac{w^2 - r_j^2}{r_j^2 w^2 - 1}
  \right)^{\rho_j}
  \prod_{k=1}^b\left(
  \frac{w^2 + s_k^2}{s_k^2 w^2 + 1}
  \right)^{\sigma_k}\times\\
\times  \prod_{l=1}^c\left(
  \frac{(w^2 - t_l^2)(w^2 -  \tbar_l^2)}
  {(t^2_l w^2 - 1)({\tbar}_l^2w^2  - 1)}
  \right)^{\tau_l}.
  \end{multline}
  Here, $\rho_j$ is $1$ or $-1$ according to whether $r_j$ is a zero
  or a pole (note that the $r_j$'s needn't be distinct), and similarly
  for $\sigma_k$ and $\tau_l$.  $n$, an odd integer, determines the
  order of the zero or pole at the origin.
  %CHANGE 
The coefficient $\epsilon$, given by $(-1)^a f(1)$, is equal to $\pm
1$ (from (\ref{eq:a)}) and (\ref{eq:c)})).
It  is straightforward to verify that
  \eqref{eq:w(z)} satisfies 
tangent boundary conditions, 
%  TBCs, 
and therefore characterises, via \eqref{eq:stereograph}, the radially
constant, conformal, tangent unit-vector fields $\nvec$ in $\octant$
for which $f$ has a meromorphic extension to $\Cc^e$.  We remark that
rational functions have been used extensively as the basis for
an ansatz for {\it skyrmions}, minimum-energy configurations of fixed
baryon number (ie, degree) of localised, nonlinear $SU(2)$-valued
fields in $\Rr^3$.  The ansatz yields very good descriptions of the
(numerically determined) energies and topologies of the true
minimisers (see, eg, \cite{hms1998}, \cite{bs2002},
\cite{ikz2004}).

%We
%  remark that the product of three (but not two) functions
%  (\ref{eq:w(z)}) is another such function.
  
  The values of the topological invariants can be computed in terms of
  the parameters of $f$.  From (\ref{eq:D}), straightforward
  calculation gives $\Dvec = \Acal\rhat$, where $\Acal$, regarded
  as a function of $w$, is given by
 \begin{equation}
   \label{eq: Acal}
    \Acal = 4|f'|^2/(1+|f|^2)^2.
 \end{equation}
 It follows that $\Omega^0 =\int_Q \Acal\, d^2 w$.  The conditions (\ref{eq: funct eqns})
above imply that $\Acal$ is invariant under $w\mapsto -w$,
 $w\mapsto \wbar$, and $w\mapsto 1/w$.  Therefore, $8\Omega^0$ is
 given by the integral of $\Acal$ over $\Cc^e$.  But this quantity
 is just ($4\pi$ times) the degree of $f$, regarded (inverse
 stereographically) as a map from $S^2$ into itself.  The degree of a
 meromorphic function is just the number of its zeros (or any other
 value) counted with multiplicity.  Therefore,
\begin{equation}
  \label{eq:Omega formula}
  \Omega^0 = \half (|n| + 2(a + b) + 4c)\pi.
\end{equation}
The edge orientations are easily determined from the values of $f$ at
$1$, $i$ and $0$ (the images of the $x$, $y$ and $z$-edges
respectively).
Let $e_x = \pm 1$ according to whether $\nvec = \pm
\xhat$ along the $x$-edge, and similarly for $e_y$ and $e_z$.
Then
\begin{equation}
  \label{eq:edge signs}
e_x = \epsilon(-1)^a,\
e_y = \epsilon(-1)^b(-1)^{(n-1)/2},\ 
e_z = \sgn\, n.
\end{equation}
The kink number $k_z$ along the $z$-face can be
computed from the change in phase of $f$ along the quarter
circle of $|w| = 1$ between $1$ and $i$.
The kink numbers $k_x$ and $k_y$ along the
$x$ and $y$ -faces
can be
computed by counting zeros of $f$ with an appropriate sign
along the respective intervals $[0,1]$
and $[0,i]$.  The result is 
\begin{eqnarray}
  \label{eq:ks}
  k_x &=& -\half (-1)^{b} e_y 
(
{\textstyle \sum_{k = 1}^b} (-1)^{k}\sigma_k + 
\half (1 - (-1)^{b}) e_z)
%\right)
,\nonumber\\
  k_y &=& -\half (-1)^a e_x 
(
{\textstyle \sum_{j = 1}^a} (-1)^{j} \rho_j + \half (1-(-1)^{a})e_z
)
,\\
k_z &=& \fourth\left(e_xe_y - n\right) 
-\half{\textstyle \sum_{j=1}^a} \rho_j
-\half{\textstyle \sum_{k=1}^b} \sigma_k
-{\textstyle \sum_{l=1}^c} \tau_l.\nonumber
\end{eqnarray}
As remarked above, the kink numbers and edge orientations
determine the trapped areas up to an integer multiple of $4\pi$.
Moreover, trapped areas with any such multiple can be realised.  For
conformal
configurations (\ref{eq:w(z)}),
it can be shown, using (\ref{eq:Omega formula}) --
(\ref{eq:ks}), that all values of the trapped area $\Omega^0$ greater than
$\Omega_{\text{min}} = 2\pi (|k_x| + |k_y| + |k_z| + \fourth)$ can be realised. 
Anticonformal configurations (obtained by replacing $w$ with $\wbar$) 
allow for all values of $\Omega^0$ less
than $-\Omega_{\text{min}}$.
%$-2\pi (|k_x| + |k_y| + |k_z| + \fourth)$.  
Further details along with a discussion of the remaining values of $\Omega^0$
%along with a discussion of the case $|\Omega^0| \le 2\pi (|k_x| + |k_y|
%+ |k_z| + \fourth)$, 
will be given elsewhere \cite{mrz2004c}.

It is possible to get better upper bounds by evaluating the integrals
in 
(\ref{eq:upper 1})
exactly.  We
do this explicitly for the configuration 
$f(w) = w$.  We call this configuration {\it unwrapped}, because it is
topologically the  simplest possible;
its kink numbers are zero and its trapped area has the
minimum allowed value.  There are in fact eight distinct unwrapped
configurations generated by the transformations $w \mapsto -w$, $1/w$ and $\wbar$.
%$, $w \mapsto 1/w$ and $w\mapsto \wbar$.  
All have the same energy.
For $f(w) = w$, it is readily computed that $\Dvec = \rvec/r^3$.
From (\ref{eq:upper 1}), the contribution 
%to (\ref{eq:upper 1}) 
to the energy from the interior face $\{x = \half L_x\}$ of $\octant$ is given by 
\begin{multline}
  \label{eq:unwrapped 1}
  8KL_x \int_0^{\hhalf L_y}\int_0^{\hhalf L_z} \frac{dy\,dz}{\ffourth
  L_x^2 + y^2 + z^2} = \\
= 2 a_{yx} a_{zx} K L_x \int_0^1 \int_0^1 
\frac{u^{-\hhalf} v^{-\hhalf}}{1 + a_{yx}^2 u + a_{zx}^2 v}\, du\,dv.
\end{multline}
The last integral may be identified with an integral representation of
the Appell hypergeometric function
$F_2(\alpha,\beta,\beta',\gamma,\gamma';s,t)$ \cite{gradshteyn}, with parameters $\alpha = 1$,
$\beta = \beta' = \half$, $\gamma = \gamma' = \smallfrac{3}{2}$ and
arguments $s = -a_{yx}^2$ and $t = -a_{zx}^2$ .  The integrals over the
two other interior faces of $\octant$ are evaluated similarly, while the
integrals over the external faces vanish by 
tangent boundary conditions.  
%TBCs.
Thus, the  unwrapped energy is
\begin{equation}
  \label{eq:unwrapped energy}
  E_0 = 8\sum_{i}  a_{ji} a_{ki} K L_i
  F_2(1,\half,\half,\smallfrac{3}{2},
\smallfrac{3}{2}, -a_{ji}^2, -a_{ki}^2),
\end{equation}
where $i$ runs over $x$, $y$ and $z$, and $(i,j,k)$ is
a cyclic ordering of $(x,y,z)$.  For a cube ( $L = K
= 1$), this gives an upper bound
of $15.3$, about 20\% more than $E_- = 4\pi$.

For conformal configurations other than the unwrapped ones, the energy
$16 K \int_{\partial R} r\Dvec\cdot {\bf dS}$ depends on the
positions $r_j$, $s_k$, $t_l$ of the zeros and poles, and therefore
can be minimised with respect to these parameters.  Since
$\int_{\partial \octant} \Dvec\cdot {\bf dS}$ is the trapped area
$\Omega^0$, and therefore is parameter-independent, it is evident that
the minimum energy is achieved by making $|\Dvec|$ small at points of
$\partial \octant$ where $r$ is large, and large at points where $r$
is small.  The local minima of $r$ are just the corners of $\octant$
on the $x$-, $y$- and $z$-edges, with projections $w = 1$, $i$ and
$0$.  For a cube, the corners are all equally close to the origin, and
the minimum energy is approached in the singular limit in which the
zeros and poles of $f$ are made to coalesce, pairwise, at $1$, $i$ or
$0$ (so that, from (\ref{eq: Acal}), $|\Dvec| = \Acal$ is made to
diverge there), leaving a single zero or pole at $w = 0$.  In this
limit, all the topologically nontrivial behaviour (kinking and
wrapping) concentrates at the edges, while away from the edges the
configuration becomes unwrapped.  This is reminiscent of the 
dipole
 configurations
of Ref.~\cite{brezis}.  However, for rectangular
prisms, the minimum energy within a conformal family may be realised
for a nonsingular configuration. This is illustrated in Fig.~\ref{fig:
  fig1}, which shows the energy (scaled by the cube root of the
volume) of the configuration $f(w) = w(w^2 + s^2)/(s^2w^2 + 1)$.  For
a cube (dashed curve), the energy approaches a minimum as $s$
approaches 1, corresponding to a configuration which is singular along
the $y$-edge.  For $L_x = 20$, $L_y = 10$, $L_z = 1$ (solid curve),
the energy has a minimum for $s$ between $0$ and $1$, corresponding to
a smooth configuration.  (The nontrivial kink number $|k_z| =1$ rules
out conformal deformations of $f$ to a configuration which is
singular along the shortest $z$-edge.)  Numerical solutions of the
Euler-Lagrange equations (\ref{eq:EL}) exhibit the same transition
\cite{mrz2004c}.
\begin{figure}
\centerline{\includegraphics[height = 2in, width = 2in]{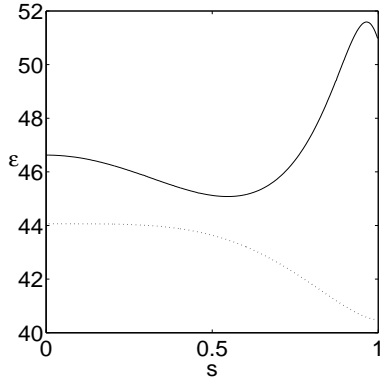}}
%\includegraphics{energy_aps.eps}
%\height = 1.5in
%\width = 1.5in
\caption{Scaled energy $\epsilon = E/V^{\tthird}$ of the conformal configuration
  $w(w^2 + s^2)/(s^2 w^2 + 1)$. Solid curve: $L_x
  = 20$, $L_y = 10$, $L_z = 1$.  Dashed curve: $L_x = L_y = L_z =
  1$. \label{fig: fig1}}
\end{figure}
One would expect to observe the smooth configurations but not
necessarily the
singular ones; in the vicinity of such an edge singularity, the liquid
crystal could melt, losing orientational order, and then relax to an
unwrapped state.

The preceding analysis indicates that the stability of topologically
nontrivial tangent director configurations in rectangular prisms
depends on the (purely geometrical) aspect ratios, a phenomenon which will be the
subject of further investigation.

\vspace{1cm}

AM was supported by an EPSRC/Hewlett-Packard Industrial CASE
Studentship.  MZ was partially supported by a grant from the Nuffield
Foundation.  We thank CJ Newton and A Geisow for stimulating our
interest in this area.

\bibliography{energy2_archive}
%\bibliography{energy2_aps}

%% \begin{thebibliography}{5}

%% \bibitem{newtonspiller} CJP Newton and TP Spiller, ``Bistable nematic
%%   liquid crystal device modelling'', 
%%   Proc. 17th IDRC (SID), 1997.

%% \bibitem{rz} JM Robbins and M Zyskin,
%%   Classification of unit-vector fields in convex polyhedra with
%%   tangent boundary conditions,  Preprint math-ph/0402025.

\end{document}